\documentclass[twocolumn,showpacs,amsfonts,aps,prc,floatfix,superscriptaddress,unsortedaddress]{revtex4-1}

\usepackage[colorlinks=true, pdfstartview=FitV, linkcolor=red, citecolor=blue, urlcolor=blue]{hyperref}
\usepackage{amsmath}
\usepackage{bm}
\usepackage{graphicx}
\usepackage{ulem}

\begin{document}

\title {
QCD equation of state with Tsallis statistics for heavy-ion collisions
}

\author{K.~Kyan}
\affiliation{Department of Mathematical and Physical Sciences, Faculty of Science, Japan Women's University,
Tokyo 112-8681, Japan}

\author{A.~Monnai}
\email[]{akihiko.monnai@oit.ac.jp}
\affiliation{Department of General Education, Faculty of Engineering, Osaka Institute of Technology, 
Osaka 535-8585, Japan}
\affiliation{Department of Mathematical and Physical Sciences, Faculty of Science, Japan Women's University,
Tokyo 112-8681, Japan}

\date{\today}

\begin{abstract}
Nonextensive statistics has attracted attention as a description of particle spectra in nuclear collisions at QCD energies.
First, we construct the equation of state by incorporating Tsallis statistics based on the hadron resonance gas and parton gas models. Thermodynamic conditions are found to impose constraints on the $q$-parameter of Tsallis distribution. Next, we apply the equation of state to the relativistic hydrodynamic modeling of nuclear collisions. The Cooper-Frye prescription is consistently modified. Numerical demonstrations indicate that the model may describe charged particle spectra at Large Hadron Collider in the transverse momentum range up to 6-8 GeV. Elliptic flow, on the other hand, suggests a narrower range of applicability. 
\end{abstract}

\pacs{25.75.-q, 25.75.Cj, 25.75.Ld}

\maketitle

\section{Introduction}
\label{sec1}
\vspace*{-2mm}

The quark-gluon plasma (QGP) \cite{Letessier:2002gp,Rafelski:2003zz,Yagi:2005yb,Wang:2016opj} created in nuclear collisions at the BNL Relativistic Heavy Ion Collider (RHIC) and CERN Large Hadron Collider (LHC) exhibits nearly-perfect fluidity \cite{Kolb:2000fha,Schenke:2010rr}. The observed hadronic spectra and flow harmonics \cite{Poskanzer:1998yz,Ollitrault:1992bk} at low momenta exhibit a clear response to the local pressure gradient unique to the hydrodynamic picture. This discovery has led to a vast number of studies aiming to understand the macroscopic evolution of the system in the hybrid modeling based on relativistic hydrodynamics and to elucidate the microscopic properties of QCD, including the equation of state and the transport coefficients. 

Hydrodynamization of the QCD matter was originally deduced as an empirical fact through comparisons of experimental data and theoretical simulations. Although no complete theoretical description has yet been available on how fluidity develops in less than $10^{-23}$ seconds, several promising models have been proposed in recent years. Notable examples include the approaches based on gauge-string correspondence \cite{Heller:2011ju} and kinetic theory \cite{Kurkela:2015qoa}, which suggest hydrodynamization before thermalization. The typical validity range of the transverse momentum for the hydrodynamic description of hadronic spectra is around 2-3 GeV. A conventional interpretation of the particle spectra at higher momenta is provided by the perturbative QCD approach. 

It has been pointed out that the nonextensive statistics, or Tsallis statistics \cite{Tsallis:1987eu,Tsallis:1999nq}, can explain the high $p_T$ power-law tail in $p$+$p$ collisions \cite{STAR:2006nmo,Tang:2008ud,PHENIX:2010qqf,PHENIX:2011rvu,ALICE:2010syw,ALICE:2012yqk,ALICE:2014hpa,ALICE:2016fzo,ATLAS:2010jvh,CMS:2010tjh,CMS:2011jlm,CMS:2011mry,CMS:2012wje,CMS:2012xvn}. The spectra in heavy-ion collisions are also considered to be described by the approach, though the presence of collective dynamics can become a non-trivial issue.
Tsallis entropy has been proposed as an extension of standard thermodynamic entropy and developed into a consistent theoretical framework. The deviation from standard thermodynamics is characterized by the deviation of the parameter $q$ from unity. There have been a large number of studies to apply the nonextensive statistics to various physical phenomena \cite{tsallis}, including relativistic nuclear collisions \cite{Wilk:1999dr,Osada:2008sw,Wilk:2008ue,Wilk:2010rr,Wilk:2012zn,Wong:2012zr,Wong:2013sca,Wilk:2013jsa,Cirto:2014sra,Wong:2015mba,Wilk:2018kvg,Biro:2003vz,Biro:2004xr,Biro:2008hz,Urmossy:2009jf,Urmossy:2012ud,Biro:2012fiy,Urmossy:2014gpa,Biro:2017arf,Biro:2020kve,Kodama:2005pp,Cleymans:2008mt,Cleymans:2011in,Cleymans:2012ya,Azmi:2014dwa,Azmi:2015xqa,Marques:2015mwa,Bhattacharyya:2015hya,Tripathy:2016hlg,Khuntia:2017ite,Bhattacharyya:2016lrk,Bhattacharyya:2017hdc,Azmi:2019irb,Marques:2012px,Alberico:2009gj,Shao:2009mu,Chinellato:2010ga,Wibig:2010si,Mishra:2013qoa,Jiang:2013gxa,Khandai:2013gva,Li:2014opa,De:2014pqa,McLerran:2015mda,De:2015hpa,Zheng:2015gaa,Thakur:2016boy,Wei:2016ihj,Gao:2016uqe,Lao:2016gxv,Parvan:2016oxk,Thakur:2016znw,Grigoryan:2017gcg,Gao:2017yas,Takacs:2019ikb,Alqahtani:2022xvo,Ahmadvand:2022urr}. 

The mechanism behind the emergence of nonextensive statistics is a topic of debate.
In the case of nuclear collisions, soft and hard sectors would not be clearly separated \cite{STAR:2006axp,Trainor:2008jp,Wong:2013sca} and thus the partonic and hadronic phase space distributions might contain strongly-interacting yet non-thermal mid-high $p_T$ component which might be associated with mini-jets. The non-extensivity in a QCD system can also be understood in the thermofractal approach \cite{Deppman:2016fxs}. The non-thermal system can be hydrodynamic, which would be consistent with the concept of hydrodynamization before thermalization.
On the other hand, most analyses simply employ a Tsallis distribution or Tsallis-extended blast wave model to fit particle spectra, which could affect the accuracy in the case of heavy-ion collisions where collective dynamics plays an essential role.
In light of the situation, it would be important to develop a hydrodynamic model based on nonextensive statistics \cite{Osada:2008sw,Takacs:2019ikb} for the quantitative argument of whether it can be justified or not as a description of relativistic heavy-ion collisions. 

In this study, we develop a Tsallis-based hydrodynamic model by taking into account the $q$-corrections to the equation of state \cite{Osada:2008sw,Alberico:2009gj,Menezes:2014wqa,Andrade:2019dgy,Deppman:2012qt} using the hadron resonance gas \cite{Dashen:1969ep} and parton gas models. Introduction of additional parameters to the hadron resonance model has been done in the case of magnetic field \cite{Endrodi:2013cs} and vorticity \cite{Fujimoto:2021xix,ExHIC-P:2020tcv}. The Cooper-Frye prescription \cite{Cooper:1974mv} is extended to ensure successful energy-momentum conservation in the particlization process. We see that thermodynamic conditions for the equation of state may limit to the parameter range of Tsallis statistics. Numerical simulations of a (2+1)-dimensional Tsallis-extended hydrodynamic model are then performed to analyze heavy-ion collisions. Charged particle spectra and elliptic flow $v_2$ are estimated in a setup similar to the situations of Pb+Pb collisions at LHC for demonstration. 

In Sec.~\ref{sec2}, we construct a QCD equation of state extended with Tsallis statistics. In Sec.~\ref{sec3}, it is embedded in the relativistic hydrodynamic model by modifying the freeze-out procedure. The model is used for the numerical estimation of particle spectra and elliptic flow in Pb+Pb collisions at different centralities. Sec.~\ref{sec4} is devoted to discussion and conclusions. The natural units $c = \hbar = k_B = 1$ and the mostly-minus Minkowski metric $g^{\mu \nu} = \mathrm{diag}(+,-,-,-)$ are used in this study.

\section{Equation of state}
\label{sec2}
\vspace*{-2mm}

We construct a nonextensive version of the QCD equation of state. The relativistic kinetic theory extended with the Tsallis distribution \cite{Tsallis:1987eu,Tsallis:1999nq} is considered for estimating the individual equations of state in the hadronic and QGP phases.

\subsection{The Model}

The macroscopic variables of the system that obey Tsallis statistics read in relativistic kinetic theory
\begin{eqnarray}
P &=& \frac{1}{3} \sum_i \int \frac{g_i d^3p}{(2\pi)^3} \frac{\mathbf{p}^2}{E_i} f_i^q , \label{eq:P} \\
\varepsilon &=& \sum_i \int \frac{g_i d^3p}{(2\pi)^3} E_i f_i^q , \label{eq:e}\\
s &=& - \sum_i \int \frac{g_i d^3p}{(2\pi)^3 E_i} [f_i^q \log_q f_i\nonumber \\
&\pm& (1\mp f_i)^q \log_q (1\mp f_i)] , \label{eq:s}
\end{eqnarray}
for the pressure, energy density, and entropy density in the limit of vanishing chemical potentials, respectively. The upper and lower signs are for fermions and bosons.

The Tsallis distribution in classical statistics is expressed as
\begin{eqnarray}
f_\mathrm{classical}(E,T,q) &=& \exp_q (-E/T) .
\end{eqnarray}
The $q$-exponential and $q$-logarithm are defined as
\begin{eqnarray}
\exp_q (x) &=& [1+(1-q)x ]^{\frac{1}{1-q}}, \\
\log_q (x) &=& \frac{x^{1-q}-1}{1-q}.
\end{eqnarray}
$i$ is the index for the particle species and $g_i$ is the degeneracy. $q$ is the parameter that characterizes the Tsallis statistics. The Boltzmann distribution can be recovered in the limit of $q=1$. See also Appendix~\ref{sec:A} for the analytical expression of the small-$(q-1)$ expansion. Quantum statistics is often introduced by substituting the regular exponential with the $q$-exponential, but this must be done with caution because $\exp_q (-E/T) \neq 1/\exp_q (E/T)$. Here we take
\begin{eqnarray}
f(E,T,q) &=& \frac{\exp_q (-E/T)}{1\pm \exp_q (-E/T)}, 
\end{eqnarray}
to allow for the convergence to the classical limit at higher temperatures and for the consistency with the results from the method of Lagrange multipliers. We use the quantum Tsallis distribution in the rest of the discussion.

It is straightforward to construct the equation of state in the hadronic phase by modifying the hadron resonance gas model, which is known to provide a reasonable description of the QCD equation of state in the thermal limit $q\to 1$. We consider the case where the deviation (q-1) is not large. It should be noted that the effect of interactions, which is encoded in resonances, does not suffer from double-counting because Tsallis statistics has its own equilibrium (q-equilibrium) and is not necessarily be a deformation from thermal statistics caused by interactions. The formulation of such theory from the grand-canonical partition function can be found, for instance, in Refs.~\cite{Deppman:2012us,Megias:2015fra}.

On the other hand, the effects of nonextensive statistics is less trivial in the QGP phase. In this study, we employ the parton gas model to embed the $q$-dependence in the QGP equation of state. 

The two equations of state are matched around the crossover temperature using the following procedure: 
\begin{eqnarray}
P(T) = P_\mathrm{had}(T) ,
\end{eqnarray}
for $T \leq T_f$ and
\begin{eqnarray}
P(T) &=& P_\mathrm{had}(T_f) + 
[P_\mathrm{QGP}(T)-P_\mathrm{had}(T_f)] \\
&\times& \{1-\exp[-c(T-T_f)]\} ,
\end{eqnarray}
for $T > T_f$ where $c$ is chosen so that
\begin{eqnarray}
\frac{\partial P}{\partial T}(T_f) = \frac{\partial P_\mathrm{had}}{\partial T} (T_f),
\end{eqnarray}
to allow smooth and continuous matching at the connection point defined with the kinetic freeze-out temperature $T_f$. It is note-worthy that instead of the conventional method where one utilizes a hyperbolic tangent function \cite{Monnai:2019hkn,Monnai:2021kgu}, we consider an exponential damping to the parton gas results at higher temperatures because (i) energy-momentum conservation must hold at freeze-out, (ii) the hadron resonance gas model may provide more reliable results than the parton gas model, and (iii) the difference between the hadronic and partonic equations of state is large for a non-zero $(q-1)$. The Cooper-Frye prescription of the kinetic freeze-out \cite{Cooper:1974mv} is based on relativistic kinetic theory and thus it provides the lowest temperature where the conversion to the hadronic picture should occur. We keep the notation $T_f$ for the connection temperature in the rest of the discussion.

\subsection{Numerical results}

We consider $u$, $d$, and $s$ quarks as components in the QGP phase and the hadronic resonances of $u$, $d$, and $s$ components with the masses below 2 GeV from the Particle Data Group list \cite{Tanabashi:2018oca} in the hadronic phase for our numerical simulation. The parameter range of $q=1$, 1.01, 1.04 and 1.07 is considered for three kinetic freeze-out temperatures $T_f=120$, 140, and 160 MeV. For each equation of state, the hadron resonance gas result is confirmed to be preserved up to $T_f$. 

\begin{figure}[tb]
\includegraphics[width=3.2in]{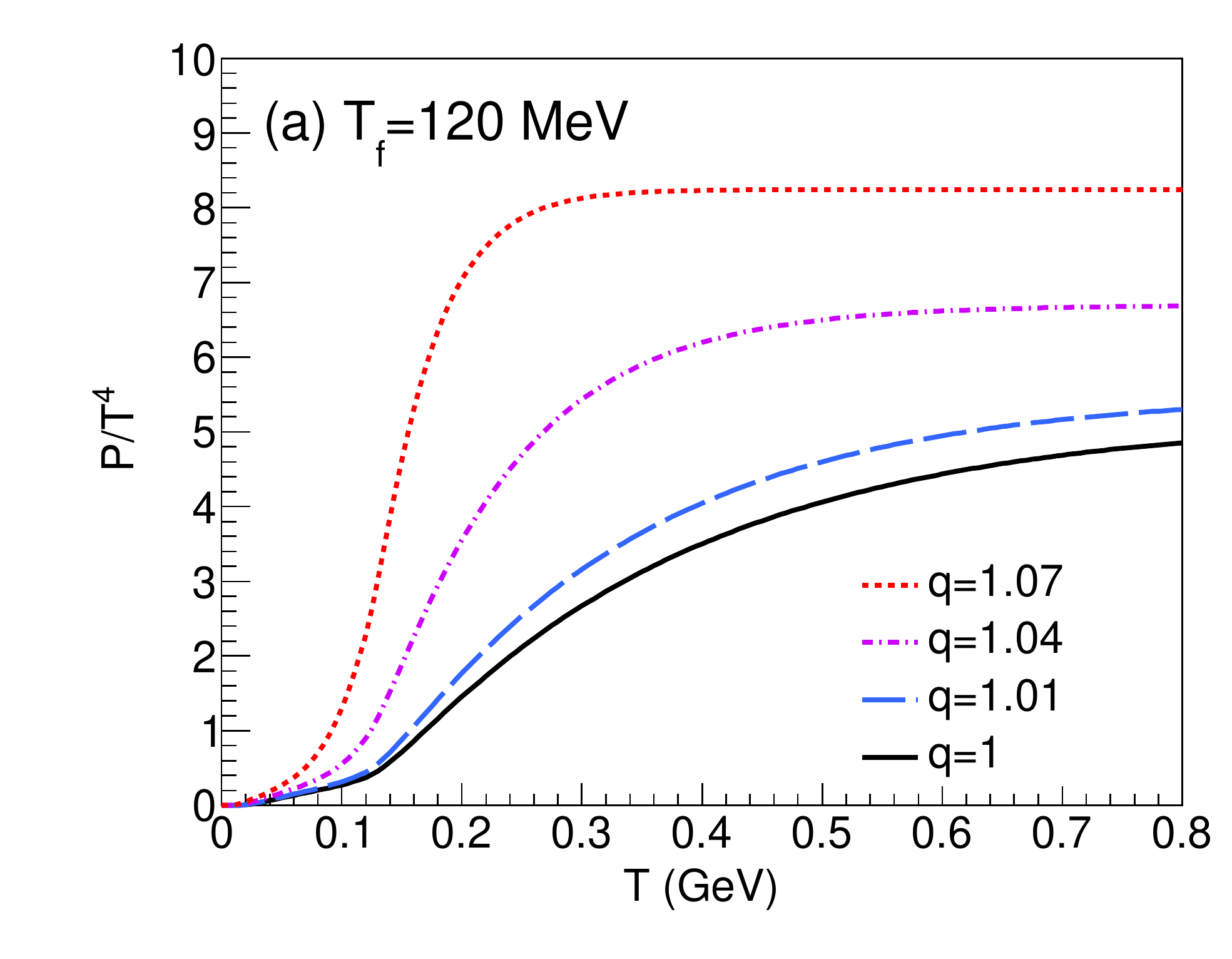}
\includegraphics[width=3.2in]{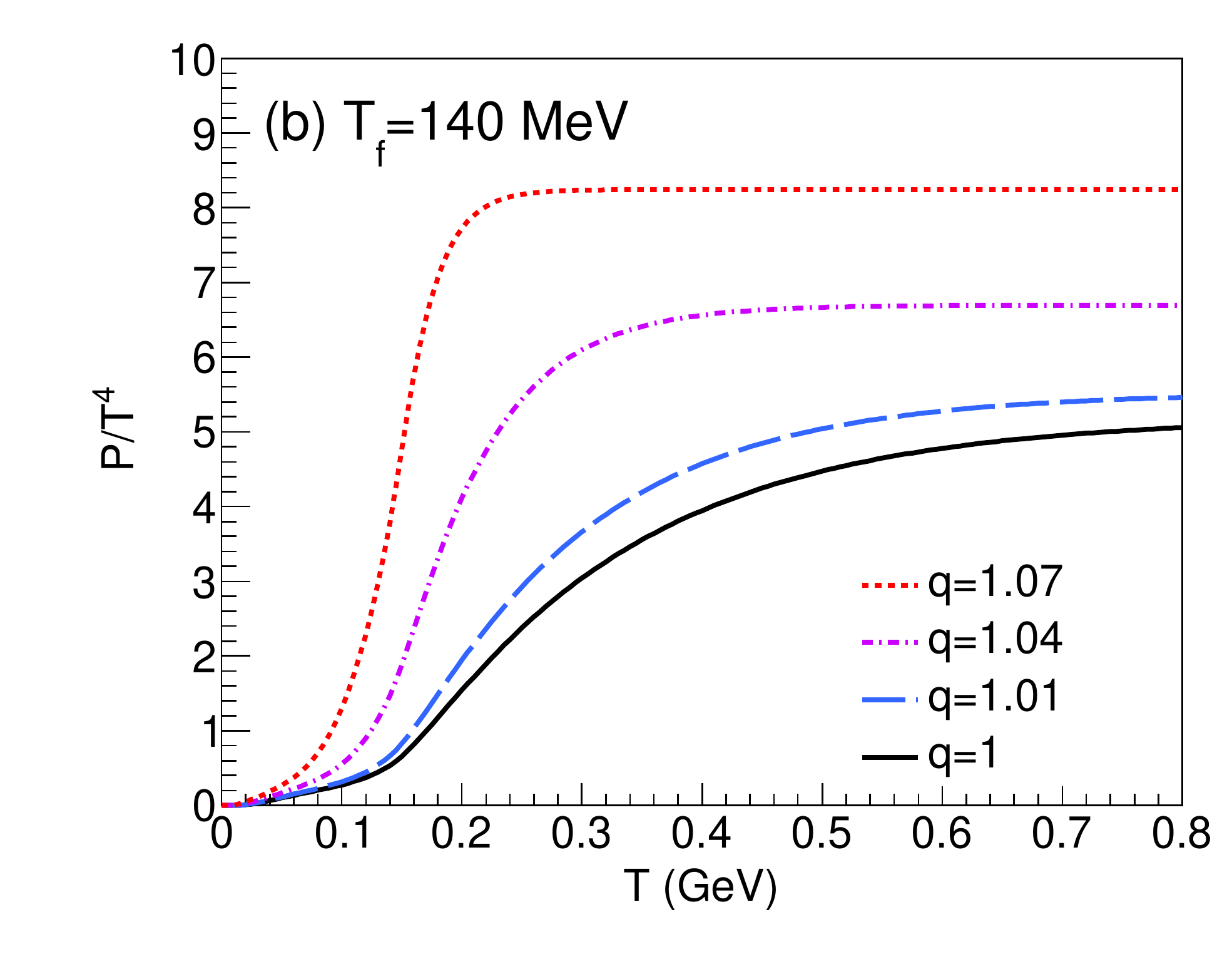}
\includegraphics[width=3.2in]{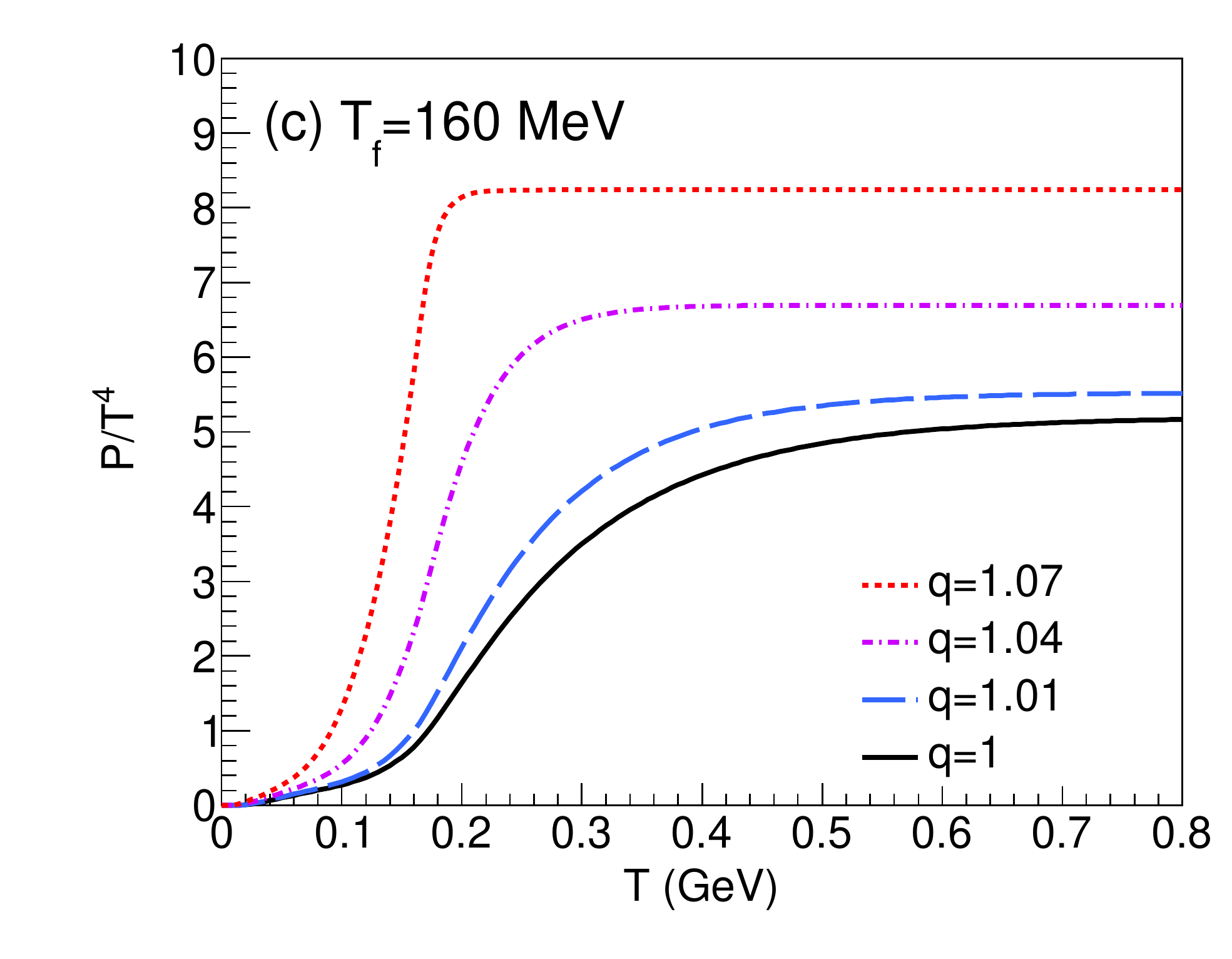}
\caption{(Color online) $P/T^4$ at $q=1$ (solid line), $q=1.01$ (dashed line), $1.04$ (dash-dotted line), and $1.07$ (dotted line) with (a) $T_f =120$, (b) 140, and (c) 160 MeV.\ \\[-11pt]
}
\label{fig:1}
\end{figure}

The dimensionless pressures $P/T^4$ as a function of temperature are plotted in Fig.~\ref{fig:1}. The pressure is sensitive to the deviation of $q$ from unity and becomes large both in the hadronic and QGP phases as $q$ increases. This can be interpreted as the effect of a larger population in large momentum regions in the Tsallis distribution for $q>1$. The hadronic pressure is found to increase faster than the QGP pressure, implying that a large $q$ may be disfavored by the thermodynamic conditions $\partial P/\partial T > 0$ and $\partial e/\partial T > 0$ around the crossover region, though the exact constraints depend on the connection procedure. It also suggests that $q$ and $T_f$ are not free parameters for the fitting of a nonextensive statistical model to experimental data. Our numerical simulations suggest that the maximum value is around $q \sim 1.07$ for $T_f$ around 140 MeV. It should be noted that the state variables do not always converge for an arbitrary $q$, imposing another constraint on the variable (see Appendix~\ref{sec:B} for details).

\begin{figure*}[tb]
\includegraphics[width=3.2in]{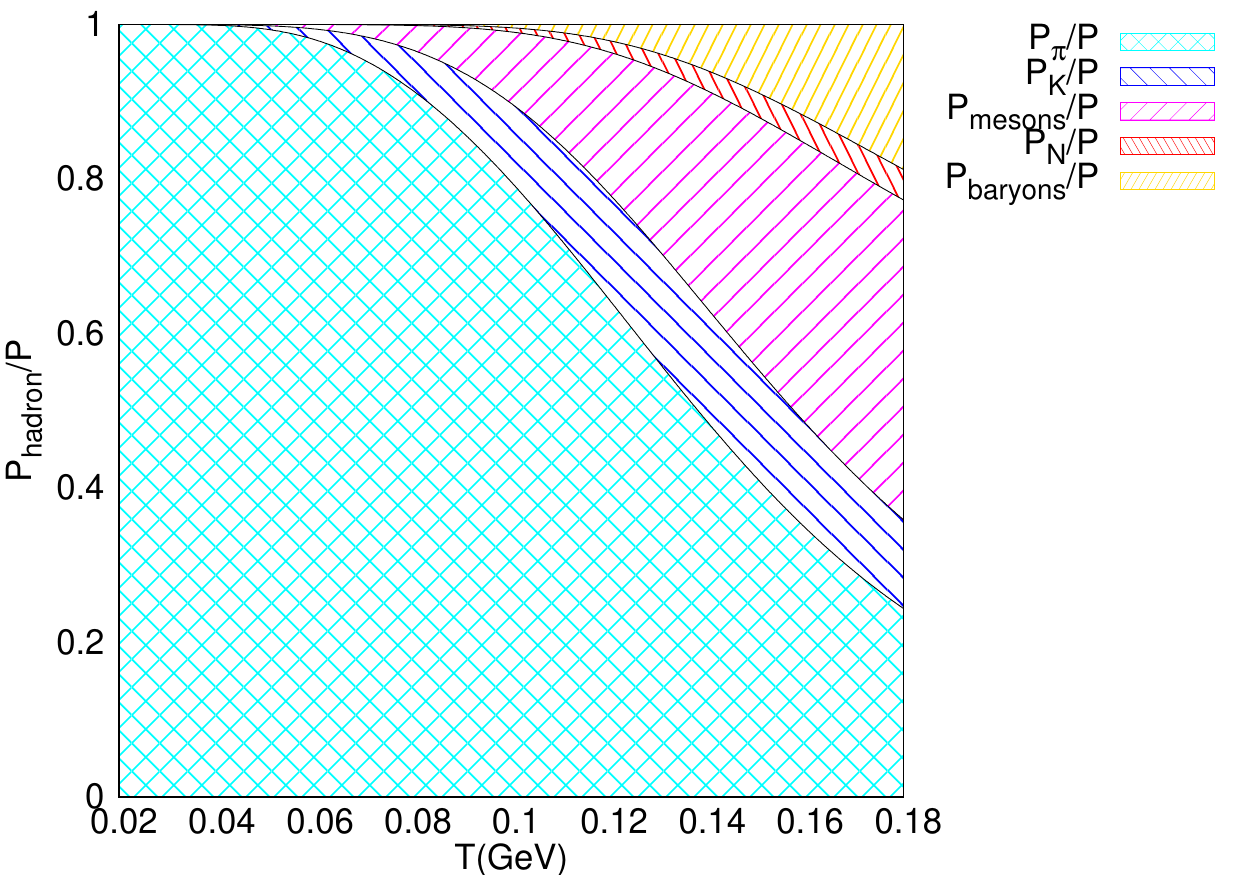}
\includegraphics[width=3.2in]{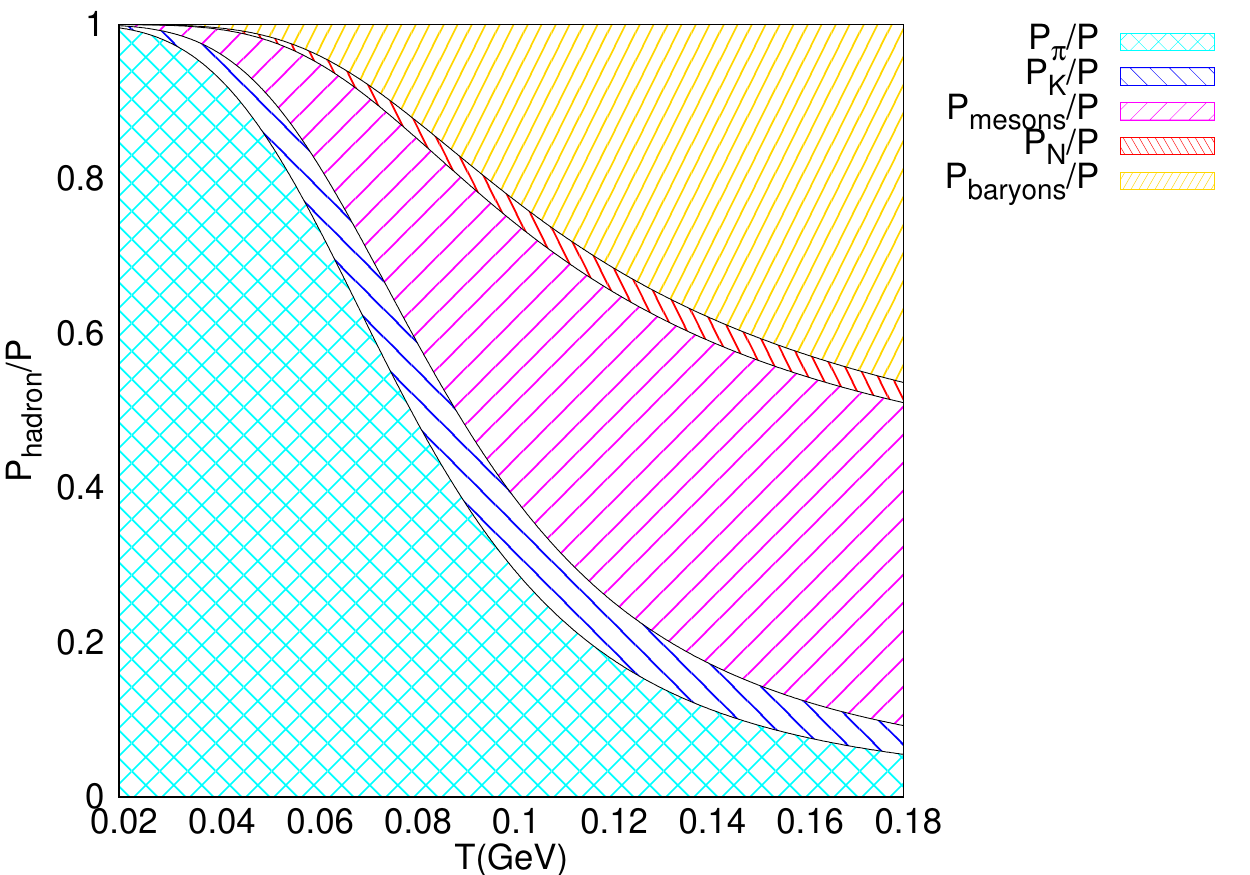}
\caption{(Color online) Ratios of the hadronic partial pressure of pions, kaons, all other mesons, nucleons, and all other baryons (from top to bottom in the legend) to the total pressure for (left) the thermal case $q=1$ and (right) a Tsallis case $q=1.07$ in the hadron resonance gas equation of state. 
}
\label{fig:2}
\end{figure*}

The ratio of the hadronic partial pressure of each component to the total pressure $P_\mathrm{hadron}/P$ is shown in Fig.~\ref{fig:2}. The thermal result ($q=1$) shows that the pion contribution is dominant at lower temperatures but is about 64.4 \% at $T = 120$ MeV, 48.6 \% at $T = 140$ MeV, and 34.8 \% at $T = 160$ MeV. On the other hand, the Tsallis result at $q = 1.07$ shows that heavier hadronic resonances have larger contributions, possibly because the effect of the Tsallis contribution masks the effect of mass difference. As a result, the relative contribution of pions is about 18.1 \% at $T = 120$ MeV, 11.5 \% at $T = 140$ MeV and 7.8 \% at $T = 160$ MeV.

\section{Application to relativistic nuclear collisions}
\label{sec3}
\vspace*{-2mm}

We develop a relativistic hydrodynamic model built consistently on the nonextensive statistics by taking into account the modifications in the equation of state and the freeze-out procedure \cite{Cooper:1974mv}. Then, we demonstrate the effect of Tsallis statistics on the observables of relativistic nuclear collisions using numerical simulations. 
 
\subsection{Equations of motion}

The hydrodynamic equations of motion are given by conservation laws and constitutive relations. In the vanishing density limit, the energy-momentum conservation $\partial_\mu T^{\mu \nu} = 0$ needs to be considered. The energy momentum-tensor with Tsallis statistics is expressed in the kinetic theory as 
\begin{eqnarray}
T^{\mu \nu} &=& \sum_i 
\int \frac{g_i d^3p}{(2\pi)^3E_i} p^\mu p^\nu f_i^q,
\end{eqnarray}
where one can see from Eqs.~(\ref{eq:P}) and (\ref{eq:e}) that the standard tensor decomposition with the flow $u^\mu$ 
\begin{eqnarray}
T^{\mu \nu} &=& (\varepsilon+P) u^\mu u^\nu - P g^{\mu \nu} ,
\end{eqnarray}
yields the state quantities in inviscid $q$-equilibrium.
The extension of the formalism to dissipative systems could involve non-triviality in the $q$-dependence of the constitutive relations as well as in the transport coefficients and will be discussed elsewhere. 
 
\subsection{Kinetic freeze-out}

The hadron number current with Tsallis statistics in kinetic theory
\begin{eqnarray}
N_i^{\mu} &=& \int \frac{g_i d^3p}{(2\pi)^3E_i} p^\mu f_i^q,
\end{eqnarray}
leads to the $q$-extended version of the Cooper-Frye formula \cite{Cooper:1974mv},
\begin{eqnarray}
E\frac{dN_i}{d^3p} &=& \frac{g_i}{(2\pi )^3} \int_\Sigma f_i^q p^\mu d\sigma_\mu ,
\end{eqnarray}
where $\Sigma$ is the freeze-out hypersurface and $d\sigma_\mu$ is the hypersurface element. The criterion for the particlization can be set using the freeze-out temperature $T_f$. It should be noted that the equation of state in the hydrodynamic evolution should match that of the hadron gas at freeze-out for successful energy-momentum conservation, which is satisfied by the extended Cooper-Frye prescription when the Tsallis equation of state is used. 

\subsection{Numerical results}

We consider Pb+Pb collisions at $\sqrt{s_{NN}}=2.76$ TeV for the situations of LHC in our demonstration. A (2+1)-dimensional hydrodynamic model \cite{Monnai:2014kqa} is used for numerical simulations.

The initial conditions are provided by the Monte-Carlo Glauber model \cite{Miller:2007ri}. There are several known variations in the modeling such as the two-component model \cite{Kharzeev:2000ph} and the quark participant model \cite{Eremin:2003qn} for the description of particle multiplicities. Hydrodynamic initial conditions have further diversity because secondary interactions modify particle distributions \cite{Kolb:2001qz}. Here, we consider the participants, which are related with soft components, as the sources.
The energy density is deposited as a Gaussian source with the width of $\sigma = 0.4$ fm at the point of each participant. 
The inelastic cross section of $pp$ collisions is $\sigma_{pp}^\mathrm{in} = 65$ mb. They are integrated for respective centrality bins before hydrodynamic simulations to obtain smooth initial conditions for simplicity in the demonstration. The normalization is determined by the data of $dN_\mathrm{ch}/d\eta$ at most central events in the corresponding experiments \cite{ALICE:2012aqc}. The hydrodynamization time is set as $0.4$ fm/$c$. Shear and bulk viscosity are turned off to elucidate the pure effect of nonextensive statistics. Its relation to the viscous corrections will be discussed in Appendix~\ref{sec:A}. Resonance decay is treated according to Ref.~\cite{Sollfrank:1990qz}.

\begin{figure}[tb]
\includegraphics[width=3.3in]{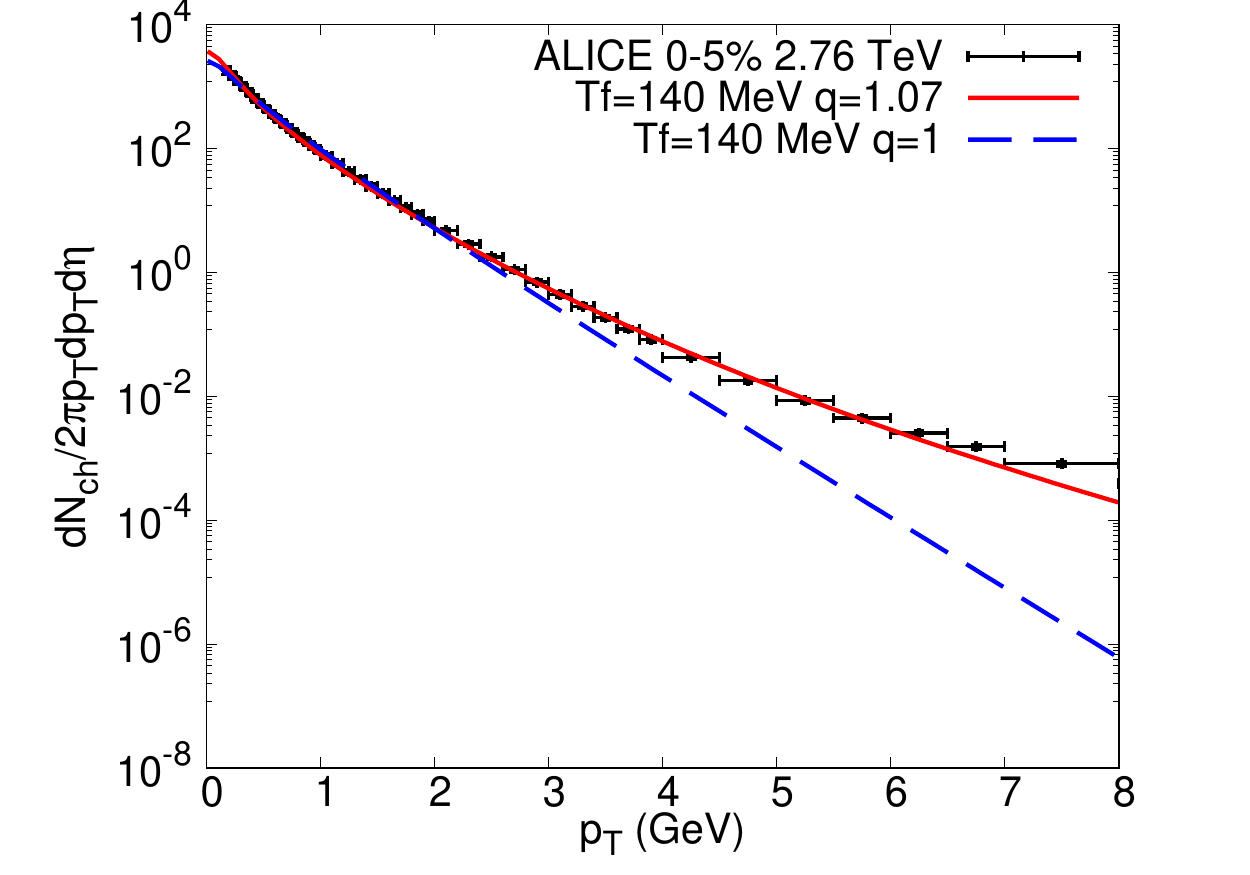} 
\includegraphics[width=3.3in]{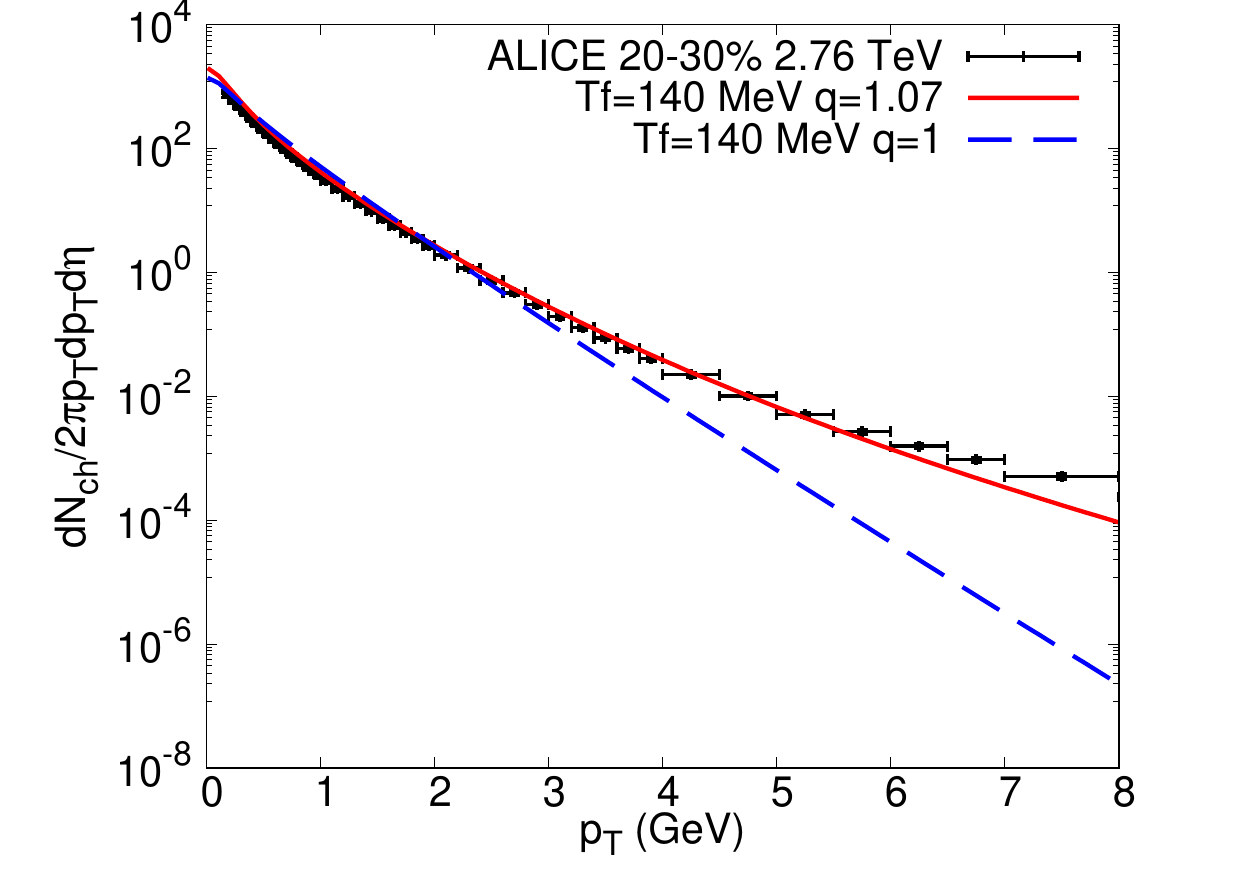} 
\caption{(Color online) $p_T$ spectra of charge particles with $q=1$ and 1.07 for (top) 0-5\% and (bottom) 20-30\% central Pb+Pb collisions at $\sqrt{s_{NN}}=2.76$ TeV compared with the ALICE data \cite{ALICE:2012aqc}.
}
\label{fig:3}
\end{figure}

The charged particle spectra of $\sqrt{s_{NN}}=2.76$ TeV Pb+Pb with $q=1$ and $q=1.07$ for the centrality classes of 0-5\% and 20-30\% are shown in Fig.~\ref{fig:3}. The experimental data from the ALICE Collaboration are plotted for comparison \cite{ALICE:2012aqc}. The kinetic freeze-out temperature is set as $T_f = 140$ MeV. The values of $q$ and $T_f$ used in the simulations are determined by the scan of parameters in the range of $1\leq q \leq 1.1$ and $0.12 \leq T_f \leq 0.16$ GeV. The traditional thermal results ($q=1$) agree with the experimental data up to 2-3 GeV while the Tsallis results ($q=1.07$) exhibit improved agreements up to 6-8 GeV in both cases in the current simulation.

The typical lifetime of the fireball is found to be shorter for a larger $q$ because the initial temperature would be smaller for a given energy density. This emphasizes the importance of developing a Tsallis-modified equation of state for a consistent framework of the nonextensive relativistic hydrodynamic model.

We also confirm that keeping the $q$ parameter consistent between the equation of state and the kinetic freeze-out is important. If the standard Cooper-Frye formula at $q=1$ is employed for the hydrodynamic simulation with the $q=1.07$ equation of state, the energy-momentum conservation is violated and the total entropy is lost, leading to reduction in particle production. Likewise, if the modified Cooper-Frye formula at $q=1.07$ is used for the hydrodynamic simulation with the thermal equation of state, the total entropy is artificially added to the system and the particle spectra is enhanced.

\begin{figure}[tb]
\includegraphics[width=3.3in]{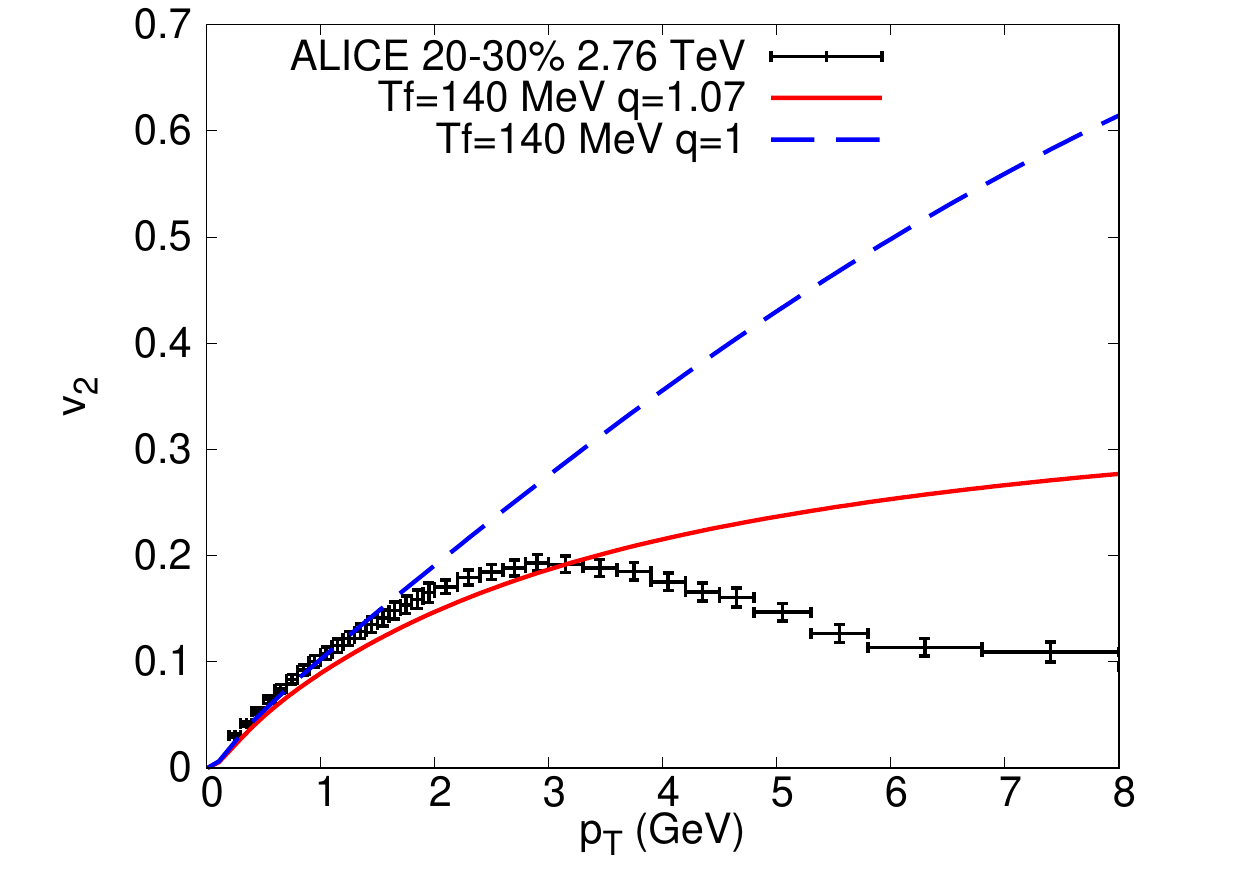} 
\caption{(Color online) Differential $v_2$ of charged particles with $q=1$ and 1.07 for (top) 0-5\% and (bottom) 20-30\% central Au+Au collisions at $\sqrt{s_{NN}}=2.76$ TeV. The corresponding ALICE data \cite{ALICE:2012vgf} are presented as a guide.
}
\label{fig:5}
\end{figure}

Next, we calculate the differential elliptic flow $v_2$ of charged particles as the normalized second harmonics of the particle spectra with respect to the event plane. The results for the 20-30\% central Pb+Pb collisions at $\sqrt{s_{NN}}=2.76$ TeV are shown in Fig.~\ref{fig:5}. It can be observed that the elliptic flow is suppressed when $q$ increases. This underdevelopment of momentum anisotropy is caused by the modified distribution at freeze-out, which effectively works as negative bulk viscosity (see Appendix~\ref{sec:A}), and the lower initial temperature for a given energy density, which leads to a shorter lifetime of the fireball. 

Because the quantity is known to be sensitive to viscosity and event-by-event fluctuations, the experimental data from ALICE Collaboration \cite{ALICE:2012vgf} are not shown for quantitative comparison but as a guide. Still, one would observe a general trend that the introduction of the Tsallis distribution reduces $v_2$, which might extend the applicability range of the hydrodynamic model from $p_T < 2$ GeV to a slightly wider range of $p_T < 4$ GeV. This was also observed in Ref.~\cite{Takacs:2019ikb}. On the other hand, the structure above $p_T > 4$ GeV might not be well reproduced, posing a challenge for the nonextensive interpretation. Possible caveats in the current analyses besides the lack of viscosity and event-by-event analyses are the fine-tuning of the freeze-out temperature and initial conditions which, might lead to a different value of $q$. 

\section{Discussion and conclusions}
\label{sec4}
\vspace*{-2mm}

We have constructed the QCD equation of state based on Tsallis statistics using the hadron resonance gas and parton gas models. The factor $q$, which characterizes the Tsallis distribution, tends to suppress the mass effect when $q-1$ increases. The pressure thus increases faster with $q$ in the hadronic phase than in the QGP phase, imposing a constraint on the maximum value of $q$ when the state variable is monotonic as a function of temperature. Several different connecting temperatures have been investigated for the matching of the hadronic and QGP Tsallis equations of state because the equation of state should be described by the hadronic one at the kinetic freeze-out temperature for successful energy-momentum conservation.

We have next embedded the obtained nonextensive equations of state in an inviscid relativistic hydrodynamic model. The Cooper-Frye prescription has been modified to make it compatible with Tsallis statistics. Numerical simulations show that a long tail structure is developed in $p_T$ spectra of charged particles for $q>1$. Compared with the standard hydrodynamic case of $q=1$, the experimental data are reproduced in a wider $p_T$ range up to $6$-$8$ GeV in the Tsallis hydrodynamic case at $T_f = 140$ MeV with $q=1.07$ in our demonstrative calculations for Pb+Pb collisions at $\sqrt{s_{NN}}=2.76$ TeV.  

The differential elliptic flow of charged particles has been found to be reduced in the Tsallis hydrodynamic simulation. It should be mentioned that na\"{i}ve quantitative comparison to the data cannot be made because the quantity is sensitive to the viscous corrections and event-by-event fluctuations. Qualitatively, on the other hand, one would observe that the extension of the applicable range of the hydrodynamic model in transverse momentum is not as large as that indicated by the particle spectra. Because the system would be strongly-coupled in a hydrodynamic description, extensive or nonextensive, sharp decrease in $v_2(p_T)$ may not be reproduced without the help of off-equilibrium corrections \cite{Teaney:2003kp,Romatschke:2007mq}.

It has been suggested that the typical value of $q$ extracted from the experimental data is around 1.1 (see, for instance, Refs.~\cite{Zheng:2015gaa,Biro:2020kve}). Our results are roughly consistent but slightly smaller compared with previous results possibly because of the full inclusion of collective dynamics. Also, a large $q$ is not favored by the thermodynamic constraint on the equation of state. 

Further future prospects include analyses of the interplay of nonextensive statistics, fluctuations and viscosity for more quantitative arguments. It would be interesting to investigate elliptic flow $v_2$ and higher-order harmonics to see whether the Tsallis scenario is consistent with the experimental data in such analyses. A hybrid model of the perturbative QCD and Tsallis hydrodynamic description at high $p_T$ may be necessary to elucidate the ratio of soft and hard components in the quark matter. Also, it would be essential to explicate the microscopic details of nonextensive statistics in the quark and hadronic matter produced in relativistic nuclear collisions.

The series of nonextensive equations of state used in the study will be made publicly available \cite{eos}.

\begin{acknowledgments}
K.K. and A.M. thank S. Yoshikawa for fruitful discussion.
The work of A.M. was supported by JSPS KAKENHI Grant Number JP19K14722.
\end{acknowledgments}

\appendix

\section{Small-$(q-1)$ expansion}
\label{sec:A}

The Tsallis distribution is expanded around the Fermi-Dirac or Bose-Einstein distribution as follows:
\begin{eqnarray}
f &=& f_\mathrm{0} + f_\mathrm{0} (1\mp f_\mathrm{0}) \frac{E^2}{2T^2} (q-1) + \mathcal{O}(q-1)^2,
\end{eqnarray}
where 
\begin{eqnarray}
f_0 &=& \frac{1}{\exp (E/T) \pm 1},
\end{eqnarray}
up to the linear order in $q-1$ in the local rest frame. The upper and lower signs correspond to fermions and bosons, respectively. It should be noted that the linear order approximation is valid when $E/T$ is not too large. The convergence of the series at higher orders will be discussed elsewhere.

It is noteworthy that the expression is similar to the bulk viscous correction to the phase-space distribution in the quadratic ansatz \cite{Dusling:2007gi}:
\begin{eqnarray}
f &=& f_\mathrm{0} + f_\mathrm{0} (1\mp f_\mathrm{0}) \frac{\mathbf{p}^2}{5T^2} \frac{\Pi}{e+P} + \mathcal{O}(\Pi)^2.
\end{eqnarray}
$\Pi$ is generally negative in an expanding system and therefore has an opposite effect on the particle spectra than the $q>1$ Tsallis statistics. Note that a full bulk viscous phase-space distribution that satisfies the Landau matching conditions is a topic of debate and may have slightly more complex expressions \cite{Monnai:2009ad,Denicol:2009am,Monnai:2010qp}.

Another difference between the expansion in $(q-1)$ and that in dissipative quantities such as $\Pi$ or $\pi^{\mu \nu}$ is that the latter has a more strict constraint that the correction should not be larger than the equilibrium pressure.

\section{Convergence of thermodynamic variables}
\label{sec:B}

The thermodynamic variables are not guaranteed to converge for an arbitrary value of $q$ in the nonextensive statistics \cite{Bhattacharyya:2016lrk}. We describe the issue in a relativistic system of massless particles in the Boltzmann limit.

One may define the moments in kinetic theory as
\begin{eqnarray}
I_{mn} &=& \frac{1}{(2n+1)!!} \sum_i \int \frac{g_i d^3p}{(2\pi)^3 E_i} p^{2n} E_i^{m-2n} f_i^q , \label{eq:Imn} 
\end{eqnarray}
for $m>2n$. The energy density and pressure are then expressed as $\varepsilon=I_{20}$ and $P=I_{21}$. The integral can be carried out analytically as
\begin{eqnarray}
I_{mn} &=& g \frac{(m+1)!}{(2n+1)!!} \frac{T^{m+2}}{2\pi^2} 
\times \frac{\Gamma[\frac{1}{q-1}-m-1]}{(q-1)^{m+1} \Gamma[\frac{1}{q-1}]} \nonumber \\
&=& g \frac{(m+1)!}{(2n+1)!!} \frac{T^{m+2}}{2\pi^2} \prod_{k=1}^{m+1} \frac{1}{1-k(q-1)}. \label{eq:Imn2} 
\end{eqnarray}
Here $g = \sum_i g_i$ is defined, which is justified in the massless Boltzmann limit because the only difference between particles is the degeneracy. The expression converges to the thermal result
\begin{eqnarray}
I_{mn} &=& g \frac{(m+1)!}{(2n+1)!!} \frac{T^{m+2}}{2\pi^2} , \label{eq:Imn3} 
\end{eqnarray}
in the limit of $q=1$.

The moment (\ref{eq:Imn2}) has poles at $q=(k+1)/k$ where $k=1,2,...,m+1$. It diverges at $q=(m+2)/(m+1)$ when $q$ increases from unity. Thus the energy density or pressure diverges at $q=4/3$ and the particle number density, expressed as $n=I_{10}$ in a single-component system, at $q=3/2$ \cite{Bhattacharyya:2016lrk}. The derivation of dissipative hydrodynamic equations can involve higher order moments up to the sixth order \cite{Monnai:2010qp}, possibly imposing a tighter constraint on the value of $q$ ($q=8/7\sim 1.14$ when $m=6$). The results reinforce the point that one should not treat $q$ as a free parameter when applying the nonextensive statistics to relativistic nuclear collisions. 

\section{Collision energy dependence }
\label{sec:C}

We explore the collision energy dependence of the Tsallis hydrodynamic description by considering heavy-ion collisions at RHIC. We consider 0-5\% and 20-30\% centrality events of $\sqrt{s_{NN}}=200$ GeV Au+Au collisions. The initial time is again set to 0.4 fm/$c$ and the freeze-out temperature to 140 MeV. The inelastic cross section of $pp$ collisions is set to $\sigma_{pp}^\mathrm{in} = 42$ mb.

\begin{figure}[tb]
\includegraphics[width=3.3in]{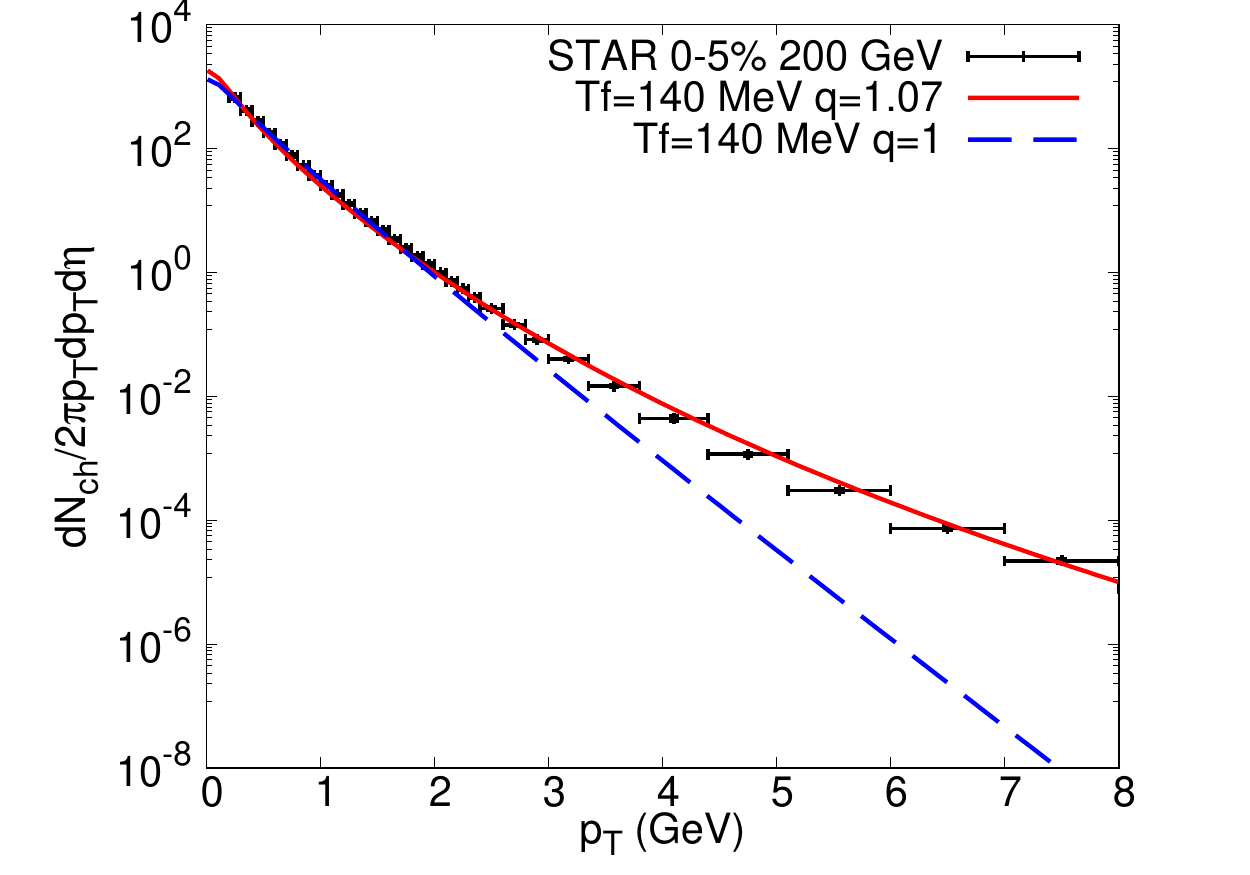} 
\includegraphics[width=3.3in]{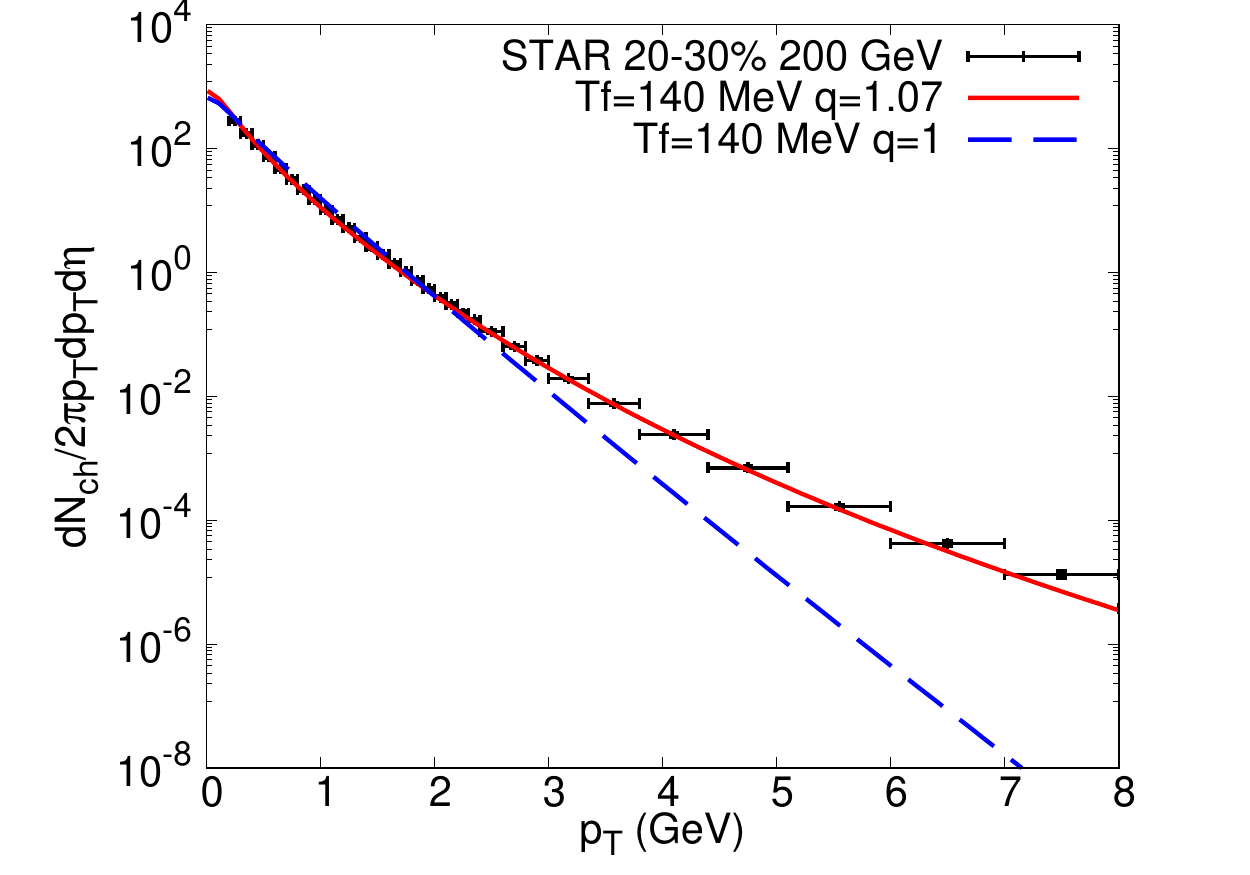} 
\caption{(Color online) $p_T$ spectra of charged particles with $q=1$ and 1.07 for (top) 0-5\% and (bottom) 20-30\% central Au+Au collisions at $\sqrt{s_{NN}}=200$ GeV compared with STAR data \cite{STAR:2003fka}.
}
\label{fig:4}
\end{figure}

The results of the numerical simulations for the charged particle spectra at $q = 1$ and $q = 1.07$ are shown in Fig.~\ref{fig:4}. The experimental data from the STAR Collaboration are plotted for comparison \cite{STAR:2003fka}. The results based on Tsallis statistics again have agreement with the corresponding experimental data in a wider momentum region up to 6-8 GeV compared with that on standard thermodynamics, which works up to 2-3 GeV. This implies that the value of $q$ might not be much affected by the collision energy.

\bibliography{tsallis.bib}

\end{document}